\documentclass{article}

\PassOptionsToPackage{numbers, compress}{natbib}

\usepackage[final]{neurips_2023}

\usepackage[utf8]{inputenc} %
\usepackage[T1]{fontenc}    %
\usepackage{hyperref}       %
\usepackage{url}            %
\usepackage{booktabs}       %
\usepackage{amsfonts}       %
\usepackage{nicefrac}       %
\usepackage{microtype}      %
\usepackage{xcolor}         %
\usepackage{graphicx}
\usepackage{adjustbox}
\usepackage{multirow}

\usepackage{amssymb}%
\usepackage{pifont}%
\usepackage{wrapfig}
\newcommand{\cmark}{\ding{51}}%
\newcommand{\xmark}{\ding{55}}%
\usepackage{subcaption}

\usepackage{algorithm}
\usepackage{algpseudocode}

\usepackage{amsmath,amsfonts,bm}

\def\eqref#1{equation~\ref{#1}}

\def\1{\bm{1}}

\def\vh{{\bm{h}}}

\def\vp{{\bm{p}}}

\def\mG{{\bm{G}}}
\def\mH{{\bm{H}}}
\def\mI{{\bm{I}}}

\def\mK{{\bm{K}}}

\def\mP{{\bm{P}}}
\def\mQ{{\bm{Q}}}
\def\mR{{\bm{R}}}

\def\mV{{\bm{V}}}
\def\mW{{\bm{W}}}
\def\mX{{\bm{X}}}

\def\mZ{{\bm{Z}}}

\DeclareMathAlphabet{\mathsfit}{\encodingdefault}{\sfdefault}{m}{sl}
\SetMathAlphabet{\mathsfit}{bold}{\encodingdefault}{\sfdefault}{bx}{n}

\def\sR{{\mathbb{R}}}

\newcommand{\softmax}{\mathrm{softmax}}

\newcommand{\Cov}{\mathrm{Cov}}

\newcommand{\proposed}{\textsc{TeNeT}}

\definecolor{LightCyan}{rgb}{0.88,0.95,1.0}

\title{Large-scale Graph Representation Learning of Dynamic Brain Connectome with Transformers}

\author{%
  Byung-Hoon~Kim\thanks{Equal contribution / $^\dagger$Corresponding author / $^\ddagger$Independent researcher.}\,\,\,$^{\dagger\,1\,2}$, Jungwon Choi$^{\ast\,3}$, EungGu Yun$^{\ddagger}$
  , Kyungsang Kim$^{2}$, Xiang Li$^{2}$, Juho Lee$^{\dagger\,3\,4}$ \\
  $^{1}$Yonsei University College of Medicine, 
  $^{2}$MGH, Harvard Medical School, $^{3}$KAIST AI, $^4$AITRICS \\
  \texttt{egyptdj@yonsei.ac.kr}, \texttt{\{jungwon.choi, eunggu.yun\}@kaist.ac.kr},\\
  \texttt{\{kkim24, xli60\}@mgh.harvard.edu}, \texttt{juholee@kaist.ac.kr}
}

\begin{document}

\maketitle

\vspace{-4mm}
\begin{abstract}
\vspace{-1mm}
Graph Transformers have recently been successful in various graph representation learning tasks, providing a number of advantages over message-passing Graph Neural Networks.
Utilizing Graph Transformers for learning the representation of the brain functional connectivity network is also gaining interest.
However, studies to date have underlooked the temporal dynamics of functional connectivity, which fluctuates over time.
Here, we propose a method for learning the representation of \emph{dynamic} functional connectivity with Graph Transformers.
Specifically, we define the connectome embedding, which holds the position, structure, and time information of the functional connectivity graph, and use Transformers to learn its representation across time.
We perform experiments with over 50,000 resting-state fMRI samples obtained from three datasets, which is the largest number of fMRI data used in studies by far.
The experimental results show that our proposed method outperforms other competitive baselines in gender classification and age regression tasks based on the functional connectivity extracted from the fMRI data.

\end{abstract}

\vspace{-4mm}

\section{Introduction}
\vspace{-1.5mm}
Functional connectivity (FC) of the brain is defined as the level of neural co-activation across time between a pair of regions, measured by neuroimaging methods such as functional magnetic resonance imaging (fMRI)~\cite{huettel2004functional}. 
Based on evidence that the pattern of FC at rest can be linked to predicting one's phenotype, interest in learning the representation of the FC has been rapidly growing with the expectation that clinical phenotypes can also be predicted~\citep{horien2022functional,morris2022predictability}.
Given the fact that FC can mathematically be regarded as a graph, graph neural networks (GNNs) have been a recent de facto choice for learning FC representations.

While researchers have witnessed promising results from the GNN-fMRI methods~\citep{bessadok2022graph}, there exist some limitations that come from the inherent structures of the model and the data.
For example, the performance of GNN models in processing FC is limited by message-passing, vulnerable to over-smoothing and over-squashing with increasing depth, and requires simplifying FC into a basic graph, thus losing some of the original rich connectivity details~\citep{rusch2023survey}.

Graph Transformers (GTs), a class of deep neural networks leveraging multi-head self-attention (MHSA), have recently shown success in various graph representation learning tasks, including in the context of functional connectivity (FC) analysis~\citep{min2022transformer,vaswani2017attention,muller2023attending}. GTs address limitations of traditional Graph Neural Networks, such as over-smoothing, by adaptively learning weights between graph components without relying on message-passing. However, challenges in effectively embedding graph data for input into Transformers remain, with recent studies focusing on improving node and edge embeddings~\citep{dwivedi2020generalization,kreuzer2021rethinking,ying2021transformers}. 
Notably, applications in FC analysis, such as those by \citet{kan2022brain} and \citet{dong2023beyond}, have demonstrated GTs' ability to encode brain graphs' structure and dynamics, offering new insights into FC from fMRI data. Yet, these approaches often overlook the temporal dynamics of FC, crucial for understanding brain function. 

A persistent issue in fMRI studies, including those utilizing machine learning methods such as GTs, is the challenge of replicability, with concerns about the generalizability of results to real-world data distributions~\citep{botvinik2022reproducibility}. While GTs have shown potential in linking fMRI signals to human phenotypes, there are lacks of evidence for their performance in external validation settings. Recent studies, however, indicate that using large-scale fMRI datasets can enhance replicability, underscoring the importance of large data volumes in fMRI research for reliable outcomes~\citep{marek2022reproducible}.

Recent studies~\citep{campbell2023dyndepnet, behrouz2023admire++, spasov2023neuroevolve, behrouz2022anomaly} have advanced the understanding of FC by focusing on its dynamic aspects and anomaly detection in brain networks, highlighting the importance of capturing temporal dynamics in FC.
However, these studies, while advancing the field in their respective areas, often do not fully address the continuous and evolving nature of FC over time, focusing more on static or snapshot-based analysis or specific aspects like anomaly detection and generative modeling.

Here, we address these issues by training and validating a novel GT-based dynamic FC representation learning method with large-scale fMRI data.
Specifically, the main goals of this work are three-fold.
One is to define the connectome embedding $\mX_{t}$ which appropriately holds the information of the brain FC at time $t$ from the raw 4D fMRI data as a combination of position, structure, and time (Section \ref{sec:embedding}).
Another is to define and train a GT $f$ such that 
$f: (\mX_{1}, \mX_{2}, ..., \mX_{T}) \rightarrow \vh_{\text{dyn}}$
where we input a sequence of connectome embeddings with $T$ timepoints and obtain the vector representation $\vh_{\text{dyn}} \in \sR ^{D}$, and $D$ is a pre-specified length to be output by the GT $f$ (Section \ref{sec:model}).
The last is to show by experiments using over 50,000 FC samples that the proposed method is capable of accurately performing classification and regression of the subject's phenotype
(Section \ref{sec:experiments}).

\begin{figure}[t]
\vspace{-5mm}
\centering
    \includegraphics[width=\textwidth]{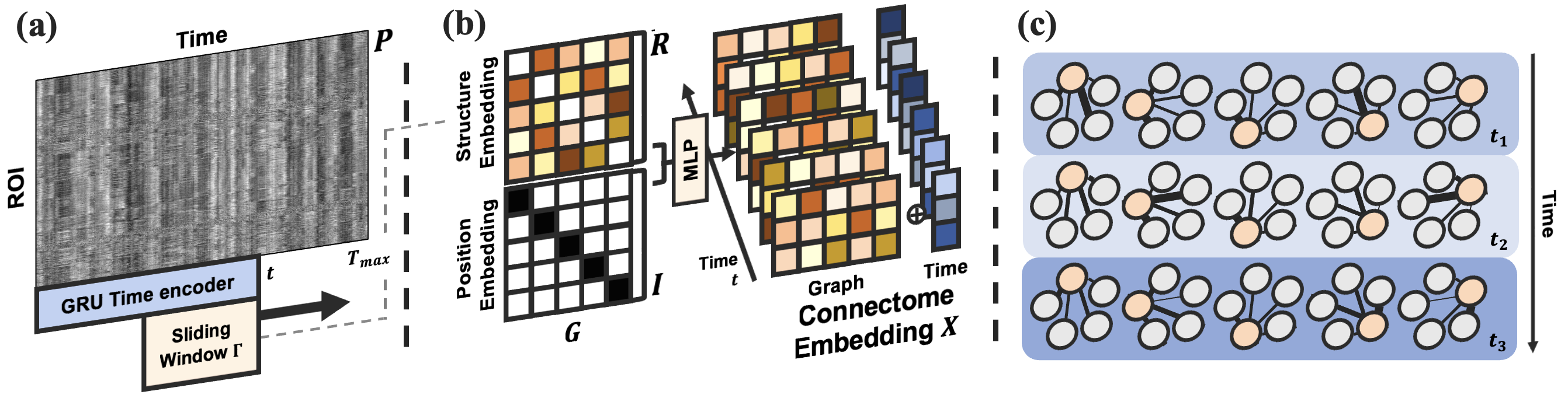}
\vspace{-4mm}
\caption{Defining the connectome embedding. (a) A GRU time encoder and the sliding-window dynamic FC approach are applied to the ROI-timeseries matrix. (b) Graph embedding $\mG$ is obtained by concatenating the structure embedding
and the position embedding
, followed by a feed-forward MLP. 
(c) The connectome embedding holds one-hop connectivity information across time at each node, which the Transformers learn self-attention weights between them.
}
\label{fig:connectome_embedding}
\vspace{-4mm}
\end{figure}

\vspace{-2mm}
\section{Main Contribution}
\vspace{-2mm}
\subsection{Defining the Connectome Embedding}
\label{sec:embedding}
\vspace{-2mm}

In defining the connectome embedding $\mX_{t}$, we start with extracting the ROI-timeseries matrix $\mP$, representing the mean BOLD signal across $N$ ROIs for $T_{\text{max}}$ timepoints. The dynamic FC graph's initial position, structure, and time are encoded using a sliding-window correlation and a GRU-based time encoding approach, following \citep{kim2020understanding,kim2021learning}. Specifically, the structure embedding $\mR_{t}$ at each time $t$ is derived from the correlation coefficients within a temporal window of length $\Gamma$, shifted over time with stride $S$, forming windowed matrices $\bar{\mP}_{t}$:
\vspace{-1.5mm}
$$
(\mR_{t})_{ij} = \frac{\Cov((\bar{\vp}_{t})_{i},(\bar{\vp}_{t})_{j})}{\sigma_{(\bar{\vp}_{t})_{i}}\sigma_{(\bar{\vp}_{t})_{j}}} \in \sR^{N \times N},
$$
\vspace{-4.5mm}

where $(\mR_{t}){ij}$ captures the edge weight between nodes $i$ and $j$. The node position is separately embedded by subtracting the identity matrix from $\mR_{t}$ to remove self-loops and then concatenating it with the identity matrix, forming $\mG := [ \; \mR_{t} - \mI \; | \; \mI \; ] \in \sR^{N \times 2N}$. This graph embedding is processed through a two-layer MLP to produce a final graph embedding in $N \times D$ dimensions.

The time embedding $\eta(t) \in \sR^D$ is the GRU output using ROI-timeseries up to the last timepoint of $\Gamma$. The final connectome embedding $\mX_{t}$ is obtained by concatenating the MLP graph embedding with the time embedding:
\vspace{-1.5mm}
$$
\mX_{t} = [\,\mathrm{MLP}(\mG)\,|\,\eta(t)\,]  \in \sR^{(N+1) \times D}.
$$
\vspace{-5mm}

This process effectively captures the one-hop connectivity information across time at each node in the FC graph.

\subsection{\proposed: Temporal Neural Transformer}
\label{sec:model}

\begin{wrapfigure}{r}{0.51\textwidth}
    \vspace{-9mm}
    \centering
        \includegraphics[width=0.51\textwidth]{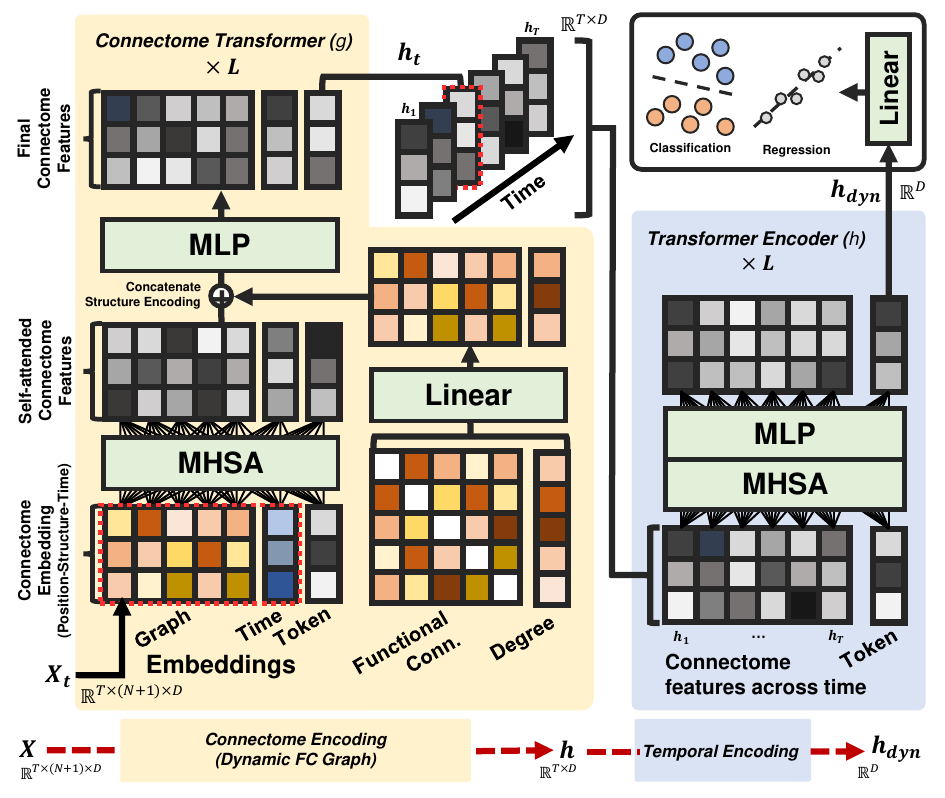}
    \caption{Schematic illustration of \proposed.}
    \label{fig:scheme}
    \vspace{-5mm}
\end{wrapfigure}

In this section, we introduce the details of our proposed method, \proposed, (Figure \ref{fig:scheme}).
To bring the learning process of the temporal information straightforwardly, we formulate the proposed method as a two-step composition of the Transformer encoders across space and time:
\begin{align}
    g&: (\mX_{1}, \mX_{2}, ..., \mX_{T}) \rightarrow (\vh_{1}, \vh_{2}, ..., \vh_{T}),  \notag\\%\quad
    h&: (\vh_{1}, \vh_{2}, ..., \vh_{T}) \rightarrow \vh_{\text{dyn}} \notag
\end{align}
where $\vh_{t}$ is a self-attended connectome feature vecture at time $t$. It can be thought that the $g$ is a self-attention across space that extracts appropriate representation at a specific timepoint, and $h$ is a self-attention across time to learn the dynamic pattern of the input fMRI signal, letting $f = h \circ g$.

As mentioned above, both $g$ and $h$ incorporate the self-attention scheme to learn the relationship between each input token within the vector-stacked matrix $\mH$ defined as:
\vspace{-7mm}

\begin{align}
    \mathrm{attention}(\mH ) &= \softmax \Bigl( \frac{\mQ \mK^{\top}}{\sqrt{D}} \Bigr) \mV,  \quad
    \mK = \mW_{\text{key}} \mH, \,\,
    \mQ = \mW_{\text{query}}  \mH, \,\,
    \mV = \mW_{\text{value}} \mH, \notag
\end{align}
\vspace{-6mm}

where $\mK$, $\mQ$, $\mV$ are transformations of input encoding to corresponding key, query, and value with learnable linear weight matrices, and $D$ is the hidden dimension. The MHSA is the self-attention parallelly projected with the multiple number of heads.

The Connectome Transformer layer processes the connectome embedding at layer $l$ using MHSA, supplemented with 1-hop connectivity $\bar{\mR}_{t}$ and degree information, then passed through an MLP for the next-layer embedding:
$
    \mZ_{t}^{l} = \mathrm{concatenate}( \{ \mathrm{attention}(\mH^{l}_t), \bar{\mR}_{t}, \Sigma^{i} (\bar{\mR}_{t})_{ij} \} ),
    \,\,\,\,
    \mH_{t}^{l+1} = \mathrm{MLP}(\mZ_{t}^{l}).
$
This design injects functional connectivity details into the Transformer, enhancing the model's depth and performance. The first layer embedding combines the connectome embedding with a random-initialized learnable token vector $\vh_{\text{token}}$, defined as $\mH^{0}_{t} := [\mX_{t} || \vh_{\text{token}} ]$. After processing through $L$ layers, the $\vh_{\text{token}}$ at each timepoint $t$ represents connectome features across time. These features are further refined by $L$ layers of a standard Transformer Encoder, culminating in a final token vector used for classification or regression tasks. For a detailed exposition of the computational process, please refer to Appendix \ref{appendix:algorithm}.

\section{Experiments}
\label{sec:experiments}

\subsection{Dataset and Experimental Setup}

\begin{table}[t]
    \caption{Summary of the experiment datasets}
    \label{tab:dataset}
    \begin{center}
    \begin{adjustbox}{width=0.8\linewidth, totalheight=\textheight, keepaspectratio}
    \begin{tabular}{llllll}
        \toprule
        Dataset & UKB & ABCD & HCP-YA & HCP-D & HCP-A \\
        \midrule
        No. Subjects & 40913 & 9111 & 1093 & 632 & 723 \\
        Gender (F/M) & 21682 / 19231 & 4370 / 4741 & 594 / 499 & 339 / 293 & 405 / 318 \\
        Age (Min-Max) & 40.0-70.0 & 8.9-11.0 & 22.0-37.0 & 8.1-21.9 & 36.0-89.8 \\ %
        \bottomrule
        
    \end{tabular}
    \end{adjustbox}
    \end{center}
    \vspace{-4mm}
\end{table}

We utilized three large-scale resting-state fMRI datasets: 1) UK Biobank (UKB)~\citep{littlejohns2020uk}, 2) Adolescent Brain Cognitive Development (ABCD)~\citep{casey2018adolescent}, and 3) Human Connectome Project (HCP)~\citep{glasser2013minimal}, each with distinct participant age groups and preprocessing protocols. For the ABCD dataset, lacking an official preprocessed version, we employed the ABCD-HCP pipeline~\footnote{\url{https://github.com/DCAN-Labs/abcd-hcp-pipeline}}. The ROI-timeseries matrix was extracted using the Schaefer atlas with 400 ROIs~\citep{schaefer2017local}. We focused on the first session of fMRI acquisition for each subject from these datasets, totaling over 50,000 samples, to mitigate sample correlation. Our experiments targeted gender classification and age regression tasks, using participant demographic data as labels. However, age regression was not applied to HCP-YA and ABCD due to limited age variability.
The experiment datasets are summarized in the Table~\ref{tab:dataset}. 
It should be noted that the datasets include over 50,000 samples in total, which is a number unprecedented in any resting-state fMRI studies by far.

Our model was structured with 4 layers ($L=4$) with each layer having a hidden dimension of 1024. 
The Adam optimizer, coupled with a one-cycle learning rate schedule, was used for optimization~\citep{smith2019super}. 
We conducted a grid search to identify the best hyperparameters, exploring batch sizes within $\{2, 4, 6, 8, 10 \}$ and learning rates within $\{ 5 \cdot 10^{-4}, 10^{-5}, 5 \cdot 10^{-6},  10^{-6}, 5 \cdot 10^{-7},  10^{-7} \}$.
Training involved 15 epochs for ABCD and UKB datasets and 30 epochs for HCP subsets, using a 5-fold cross-validation method. All experiments were executed on an NVIDIA GeForce RTX 3090.

\subsection{Comparative Experiment}

\begin{table}[t]

\caption{Performance table on our benchmark datasets.}
\label{tab:performance}
\begin{center}
\begin{adjustbox}{width=0.8\linewidth, totalheight=\textheight, keepaspectratio}
\begin{tabular}{lllllllll}
    \toprule
    Dataset & HCP-YA & \multicolumn{2}{l}{HCP-D}& \multicolumn{2}{l}{HCP-A} & \multicolumn{2}{l}{UKB} & ABCD \\
    \cmidrule(r){1-1}
    \cmidrule(r){2-2}
    \cmidrule(r){3-4}
    \cmidrule(r){5-6}
    \cmidrule(r){7-8}
    \cmidrule(r){9-9}
    Feature & Gender  &  Gender & Age & Gender & Age & Gender & Age & Gender \\
    \midrule
    \proposed  & 95.07  & \textbf{82.89} & \textbf{0.6878} & \textbf{89.05} & \textbf{0.6663} & 98.37 & \textbf{0.4768} & \textbf{90.21} \\ 
    BNT        & 95.49  & 81.29 & 0.6756  & 86.63 & 0.5828  & \textbf{99.04} & 0.4635 & 85.98  \\
    STAGIN     & \textbf{95.50} & 70.36 & 0.3966 & 84.57 & 0.4473 &  98.61 & 0.4047  & 81.61   \\
    GIN & 86.49  &  62.86 & 0.3054 & 68.91 & 0.3130 & 96.67 & 0.3904 & 73.19 \\ 
    \bottomrule
    \multicolumn{9}{l}{
    \begin{minipage}[t]{13cm}
    * Performance of gender classification and age regression are reported with AUROC (\%) and $R^2$ scores, respectively.
    \end{minipage} 
    }

\end{tabular}
\end{adjustbox}
\vspace{-7mm}
\end{center}
\end{table}

The performance of \proposed\,is validated by comparing it with several baseline methods on our benchmark datasets.
The baseline methods include the BNT~\citep{kan2022brain}, a GT-based static FC method, STAGIN~\citep{kim2021learning}, a GNN-based dynamic FC method, and GIN~\citep{kim2020understanding}, a GNN-based static FC method.
Performance of gender classification and age regression is evaluated with the area under the receiver operating curve (AUROC) and the $R^{2}$ scores, respectively.

The main results are summarized in the Table~\ref{tab:performance}.
From the comparative experiments, it can be seen that \proposed outperforms other baseline methods in most of the phenotype prediction tasks.
We expect that this gain in performance comes from the ability of \proposed to exploit dynamic information of the FC that changes over time.

\subsection{Ablation Study}
\vspace{-2mm}
\begin{figure}[h]
\begin{minipage}{\textwidth}

  \begin{minipage}[h]{0.48\textwidth}
    \captionsetup{type=table}
    \caption{Ablation results evaluating the impact of temporal information.}\label{tab:ablation}
    \centering
    \begin{adjustbox}{width=\linewidth, totalheight=\textheight, keepaspectratio}
    \begin{tabular}{c|c|c}
    \toprule
    Use time encoding & Dynamic graph & AUROC \\
    \midrule
    \cmark & \cmark  & 89.82 \\
    \xmark & \cmark & 88.45 \\
    \cmark & \xmark & 84.57 \\
    \bottomrule
    \end{tabular}
    \end{adjustbox}
  \end{minipage}
  \hfill
  \begin{minipage}[h]{0.51\textwidth}
    
    \centering
    \scalebox{1.0}[0.9]{
        \includegraphics[width=\textwidth]{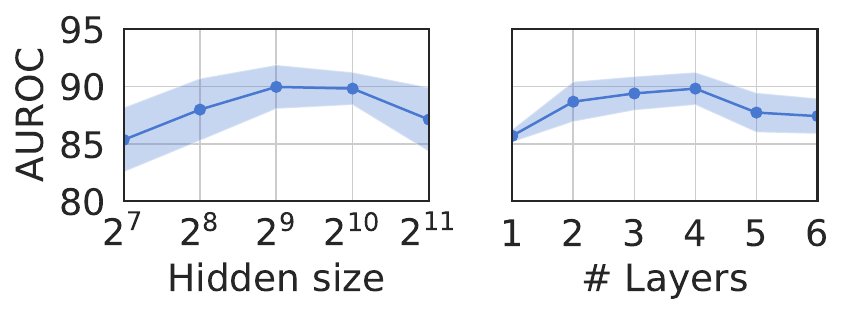}
    }
    \vspace{-5mm}
    \caption{Ablation results on model size of {\proposed}.}
    \label{fig:ablation}
  \end{minipage}
\end{minipage}
\vspace{-2mm}
\end{figure}

We conducted an ablation study on \proposed\,to assess the impact of temporal information, which involved two scenarios: 1) removing the GRU-derived time embedding and 2) replacing the dynamic graph embedding with a static one. The results, as detailed in Table \ref{tab:ablation}, indicate a decrease in performance for both scenarios in the HCP-A gender classification task, demonstrating the importance of temporal information in our model.  Additionally, sensitivity tests for the hidden dimension size and the number of layers revealed optimal performance at specific thresholds, with diminishing returns beyond these points as shown in Figure \ref{fig:ablation}.
\vspace{-2mm}

\section{Conclusion}
We propose \proposed, a GT-based method for learning dynamic FC of the brain with position, structure, and time embeddings. 
Experiments with large-scale resting-state fMRI datasets confirm the validity of \proposed.
Further studies of \proposed on attention interpretability, external validation performance, and theoretical understanding are expected to provide valuable insight into applying GT to the time-varying FC graph.

\begin{ack}
This work was partly supported by Basic Science Research Program through the National Research Foundation of Korea(NRF) funded by the Ministry of Education (NRF-2022R1I1A1A01069589), the National Research Foundation of Korea(NRF) grant funded by the Korea government(MSIT) (NRF-2021M3E5D9025030) and Institute of Information \& communications Technology Planning \& Evaluation (IITP) grant funded by the Korea government(MSIT) (No.2019-0-00075, Artificial Intelligence Graduate School Program (KAIST)).
\end{ack}

\clearpage
\newpage
\appendix
\section{Detailed Algorithmic Description of \proposed}
\label{appendix:algorithm}

We provide a detailed description of the \proposed's computational process. The following pseudocode outlines the model's core algorithmic steps, delineating both spatial and temporal attention mechanisms within the Connectome Transformer and Transformer Encoder modules and complements Figure 2 in the main manuscript, providing a more comprehensive understanding of how the both of modules are integrated in processing pipeline of \proposed.

\begin{algorithm}
\caption{Algorithmic Flow of \proposed}
\begin{algorithmic}[1]
\State \textbf{Input:} Time-sequenced connectome embeddings $\mX_{t}$ for $t \in \{1, \ldots, T\}$, where $T$ is the total number of timepoints, and Learnable token vector $\vh_{\text{token}}$.
\State \textbf{Output:} Final token vector $\vh_{\text{dyn}}$ 

\Procedure{\text{Connectome Transformer\,}}{$g$}
    \For{$t = 1$ \textbf{to} $T$}
        \State $H^{0}_{t} \gets \mathrm{Concatenate}(\mX_{t}, \vh_{\text{token}})$ \Comment{Initial embedding for time $t$}
        \State $\bar{\mR}_{t}  \gets \mathrm{Linear}(\mR_{t})$ \Comment{Linear transformation of structure encoding}
        \State $\Sigma^{i} (\bar{\mR}_{t})_{ij} \gets  \mathrm{Linear}(\Sigma^{i} (\mR_{t})_{ij})$ \Comment{Node degree information}
        \For {$l = 1$ \textbf{to} $L$}  \Comment{$L$ is the number of Transformer layers}
            \State  $\mZ_{t}^{l} \gets \mathrm{Concatenate}( \{ \mathrm{attention}(\mH^{l}_t), \bar{\mR}_{t}, \Sigma^{i} (\bar{\mR}_{t})_{ij} \})$ \Comment{Spatial Attenttion}
            \State $\mH_{t}^{l+1} \gets \mathrm{MLP}(\mZ_{t}^{l})$
    \EndFor
    \State $\vh_t \gets \mH_{t}^{L}[\texttt{token}]$  \Comment{Extract token vector as connectome feature}
    \EndFor
    \State $\vh \gets \mathrm{Concatenate}(\vh_1, \vh_2, ..., \vh_T)$
    \State \Return $\vh$
\EndProcedure

\Procedure{\text{Transformer Encoder\,}}{$h$}
    \State $H^{0} \gets \mathrm{Concatenate}(\vh, \vh_{\text{token}})$ \Comment{Initial embedding}
    \For {$l = 1$ \textbf{to} $L$} 
        \State $\mZ^{l} \gets \mathrm{attention}(\mH^{l})$ \Comment{Temporal Attenttion}
        \State $\mH^{l+1} \gets \mathrm{MLP}(\mZ^{l})$
    \EndFor
    \State $\vh_{dyn} \gets \mH^{L}[\texttt{token}]$ \Comment{Extract token vector as final token vector}
    \State \Return $\vh_{\text{dyn}}$
    
\EndProcedure
\end{algorithmic}
\end{algorithm}

\end{document}